# A condensed-matter analogue of the false vacuum

Mark Gibbons CEng MEI

Target Carbon Limited, 271 Coppice Road, Poynton, Stockport, Cheshire, SK12 1SP, United Kingdom

Email: markgibbons@targetcarbon.co.uk

Abstract

Through experimental investigation into the behaviour of a polar dielectric working fluid, an ideal 'quasi-thermodynamic' cycle has been established. Particular stages of this cycle are described in terms of a condensed-matter analogue of the false vacuum, when operating under negative pressure. The cycle is established between 37˚C and 15˚C under isochoric conditions. Phase-change work is created in two-directions, positive expansion-work and negative contraction-work. A large proportion of the expansion-work derives from a cooling process where the fluid exhibits negative heat capacity. When heat flux ceases, the fluid becomes unstable and heat capacity switches from negative to positive, displaying a 'non-equivalence of ensembles' phase-change.

Whilst elements of the fluid behaviour can only be described by the statistical mechanics of non-equilibrium systems, the calculated equations of state for classical thermodynamics are confirmed to be accurate from the experimental investigation. However, the classical thermodynamic calculations for cycle-efficiency do not produce a symmetry of energy conservation. This suggests that an additional form of energy, having long-range interaction and distinct from heat and work input, is involved in the performance of the quasi-thermodynamic cycle. The expansion of a negative pressure fluid that contains inclusion compounds appears responsible for this potential energy interaction as an analogue of the false vacuum potential that can be explained by application of the virial theorem.

Introduction

The modelling of physical processes through quasi-static and isolated-system assumptions has underpinned the development of classical thermodynamics as a foundational topic within science from the 19$^{th}$ century. However, since the beginnings of thermodynamics as an academic discipline, and its subsequent codification into physical laws, it has been recognised that the quasi-static and isolated-system assumptions cannot fully represent complex, real-world situations.

Rankine identified that the properties of saturated steam could not be fully described within a closed system and that its apparent specific heat could be negative, a condition forbidden by classical thermodynamic theory [1]. In addition, the virial theorem of Clausius [2] produces a general equation relating the time-average of the kinetic energy of a stable system bound by potential forces with that of the total potential energy of the system. The virial theorem holds for systems not in thermal equilibrium and its application can also result in negative heat capacity. The situations described by Rankine and Clausius result in non-equilibrium, long-range interactions, without involving the transfer of matter. From Carathéodory's principle, it follows that such systems cannot be isolated in order to comply with the second law of thermodynamics [3].

Due to the complexity of the phenomena arising from our experimental investigations, it has been necessary to move beyond classical thermodynamics to find solutions in statistical mechanics and the physics of non-equilibrium systems. It has also been necessary to validate our unusual results in terms of thermodynamic processes and phenomena generally ascribed to the fields of cosmology and chemistry. Within statistical mechanics, it is well known that the negative specific heat region of the micro-canonical distribution can be replaced by a phase transition in the canonical distribution [4,5]. Although this correspondence is well established, the author is unaware of any physical experiment previously capable of measuring the discrepancy between the micro-canonical and canonical macrostate properties of a system. The situation has been described by Touchette *et al* with the conclusion that any value determined for non-equivalence can only be derived within the latent heat range of the system being considered [6].

The statistical mechanics of systems dominated by gravity has close connections with areas of condensed matter physics and fluid mechanics [7,8]. Furthermore, gravitating systems in virial equilibrium are similar to normal systems with short-range forces on the verge of a phase transition [4,5]. For the exposition of our experimental results revealing the creation of a quasi-thermodynamic cycle based upon a polar dielectric working fluid operating under non-equilibrium conditions, the physics of the metastable false vacuum are considered. Metastable scalar fields can exist in a false vacuum state at a local minimum of energy. Additionally, the unusual properties of the false vacuum arise from its large negative pressure.

In Guth's model of inflation, as the energy of matter increases by a factor of $10^{75}$ or more, the energy of the gravitational field becomes more and more negative to compensate. The total energy, matter plus gravitational, remains constant and very small, and could even be exactly zero [9]. The existence of comparable kinetic and potential energies is a feature of self-gravitating systems that are in virial equilibrium and can be understood in terms of the constant energy attribute of the micro-canonical ensemble [4]. However, since the ultimate source of the gravitational potential is unknown, it is not clear that the system is strictly isolated.

Non-equilibrium, long-range interactions within matter and the gravitational field of the false vacuum both emerge as counter-acting, restorative forces, reminiscent of Anaximander's *apeiron* or the *palintropos* of Heraclitus. The negative energy potential of these equal-and-opposite forces forms the basis of the analogy presented here between a metastable working fluid and the cosmological false vacuum. The 'borrowing' of energy from the gravitational field also finds a counterpart in Onsager reciprocal relations and dissipative structures for open, irreversible, non-equilibrium systems [10,11]. As complex structures are emergent within self-gravitating systems, so far-from-equilibrium systems can give rise to regularities and symmetries in the form of organised, dissipative structures [12].

Non-additivity is a feature of non-equilibrium, self-gravitating systems with *a priori* long-range interactions [13]. This non-additivity of thermodynamic potentials and energy frequently leads to a 'non-equivalence of ensembles' that can in turn give rise to negative heat capacities. Negative heat capacities in the micro-canonical ensemble correspond to phase-transitions in the canonical ensemble [14,15,16]. These phase transitions may exhibit peculiar behaviour such that a system can jump from a lower free energy level to a higher one [17].

Systems with long-range interactions are characterised by an interaction potential (*V*) that decays with inter-particle distance with an exponent smaller than the number of dimensions (*d*) of the embedding space, ie. $V(r) \sim 1/r^\alpha$, with $\alpha \leq d$. The internal energy of such systems lacks extensivity and additivity. Although the extensivity can be restored by scaling the interaction potential with the number of particles, the non-additivity still remains [15,18].

The fluctuation theorem [19] holds for isolated, non-extensive systems, which are systems with long-range interactions. It predicts that, under very general assumptions, stable thermal-equilibrium configurations of isolated systems (micro-canonical ensembles) near instability always have negative heat capacities, which switch to be positive when systems become unstable. These results imply that a stable isolated system (a micro-canonical ensemble) near instability is necessarily unstable when placed in a heat bath (a canonical ensemble) resulting in a 'non-equivalence of ensembles' [20].

The experimental investigation of an unusual polar dielectric fluid of our own formulation uncovered the peculiar behaviours of negative heat capacity and spontaneous nucleation. Further research revealed these behaviours to be characteristic of metastable, non-equilibrium and self-gravitating systems. We describe how these conditions can arise in the polar dielectric fluid and how the associated long-range interactions can be incorporated into an ideal 'quasi-thermodynamic' cycle.

The findings of our experimental investigations quantify the discrepancy between the micro-canonical and canonical treatment of a system governed by long-range, van der Waals interactions. We have been able to establish a 'quasi-thermodynamic' cycle that incorporates long-range potential energy interactions arising from a non-equilibrium, negative pressure, metastable system created with a polar dielectric fluid of our own formulation. This is possible since in satisfying the symmetry of energy conservation, all thermodynamic and non-equilibrium energies must sum to zero when proceeding through a full cycle.

This research also seeks to establish that when our polar dielectric fluid formulation is subjected to negative pressure, a decrease in phase density leads to the formation of new, lower density, dissipative structures driven by excess negative entropy [21]. Through experimental investigations into the irreversible behaviour of this negative pressure, non-equilibrium fluid, evidence is sought for a long-range energy potential that, being associated with negative pressure expansion, can be considered an analogue of the false vacuum. We seek to incorporate the excess negative entropy and long-range interactions into a quasi-thermodynamic cycle where the sum of changes in energy and thermodynamic potentials, plus any non-equilibrium energy potentials, must satisfy the symmetry of energy conservation when completing a quasi-thermodynamic cycle.

Methods and materials

The original objective for this research was to reduce compression-stage losses associated with thermodynamic cycles. Isochoric thermo-compression [22] of polar dielectric fluids at temperatures below 100˚C was selected as a suitable area of investigation to reduce the mechanical compression penalties associated with thermodynamic cycles.

Our initial modelling with the NIST REFPROP program/ database [23] together with the experimental investigation targeted thermal instabilities arising from strong interactions between attractive and repulsive molecular forces as a source of thermo-compression energy. Through empirical observation and results, it became possible for us to establish a repeatable quasi-thermodynamic cycle through the controlled application of heat flux. With the application of heating and cooling only (ie. no external work input) an ideal, isochoric cycle has been established. The refined process arrived at is examined in detail below.

The polar dielectric working fluid is a multi-component formulation, consisting of water and methane within an inhibitor solvent engineered such that the equilibrium between attractive and repulsive molecular forces is readily destabilised when the fluid is subjected to heat flux. The nature and purpose of the clathrate inhibitor has previously been described in detail by Perrin *et al* [24]. Our own inhibitor formulation is designed to maximise the phase-change work of a quasi-thermodynamic cycle at low-temperatures (ie. below 40˚C). Changes in the polar dielectric properties are thermally induced with a view to creating a quasi-thermodynamic cycle through non-equilibrium, irreversible expansion and contraction.

Negative pressure conditions are created by means broadly similar to the Berthelot-method [25]. The polar dielectric working fluid, as specially formulated by us, fills a previously evacuated stainless-steel sample vessel (50ml), as **Fig. 1** below. The initial fill-pressure is chosen such that the fluid will exhibit negative pressure behaviour under certain heat flux conditions, as described later. This is achieved through close control of fluid dosage together with pressure (bottled nitrogen) and vacuum (generated) indirectly applied via the 5 litre bladder pressure vessel to produce the required initial vapour pressure for the volatile fluid mixture. A fluid with negative pressure is able to resist an applied tension to produce a negative internal pressure. Such a possibility is predicted by the van der Waals equation of state where negative pressure is seen as a property of metastable fluids [26].

The sample vessel is sealed-off using isolation valves. It is completely immersed in a relatively large heat bath (70 litres) where the temperature of the bath is varied with an electric element and a refrigeration dip cooler. The temperature and pressure of the working fluid are measured at five-second intervals with sensors that are in direct contact with the fluid and recorded by a PLC/ PC monitoring system. We subject the working fluid to a sequence of heating and cooling operations over an 11-hour cycle period in order to produce maximum pressure-volume (*P-v*) work from low-temperature conditions within the large heat bath. This timeclock-controlled sequence has been optimised for the particular apparatus size (see **Fig. 1**) and based upon a large number of empirical results.

The sequence of heating and cooling is as follows:

    Stage 1-2: heating with electric element

    Stage 2-3: continued heating with electric element

    Stage 3-4: cooling with refrigeration dip cooler

    Stage 4-1: removal heat flux, electric element and dip cooler off

The temperature and pressure measurements recorded are entered into the NIST REFPROP program/ database [23], which calculates the thermodynamic properties from equations of state. The calculations are in accordance with GERG-2008 modified by the Kunz and Wagner Model 0 (KW0) [27]. There is no external work input during any stage of the cycle and the dielectric constant of the fluid is not measured.

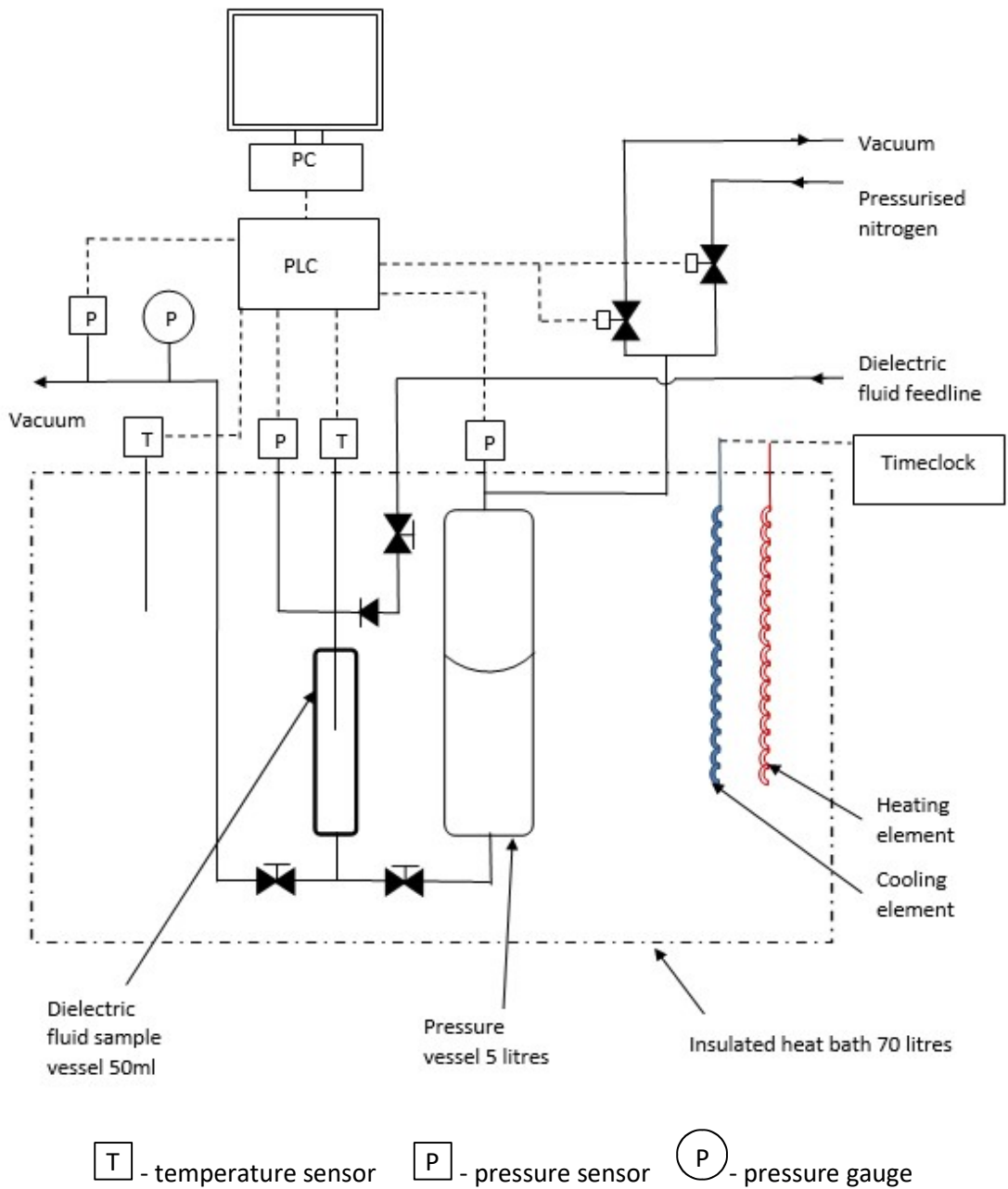

Fig. 1: Schematic arrangement of the experimental apparatus

Results

**Table 1** gives a summary of the experimental results and is included in Appendix 1. All values for energy and thermodynamic potentials are derived from the pressure and temperature measurements by the REFPROP program/ database. The liquid and vapour saturation curves, isothermals, and isobars determined by this model are confirmed to be accurate through empirical experimental measurements.

The following charts are derived from these results. Analysis and discussion of the results follow in the next section. The charts (except Fig. 8) plot the quasi-thermodynamic results from the 5-second interval data over the 11-hour cycle. Distinctive phase-changes are identified at Points 1, 2, 3, 3a and 4.

**Fig. 2** reveals a distortion of short-range energy interactions through the presence of a non-concave entropy function of internal energy:

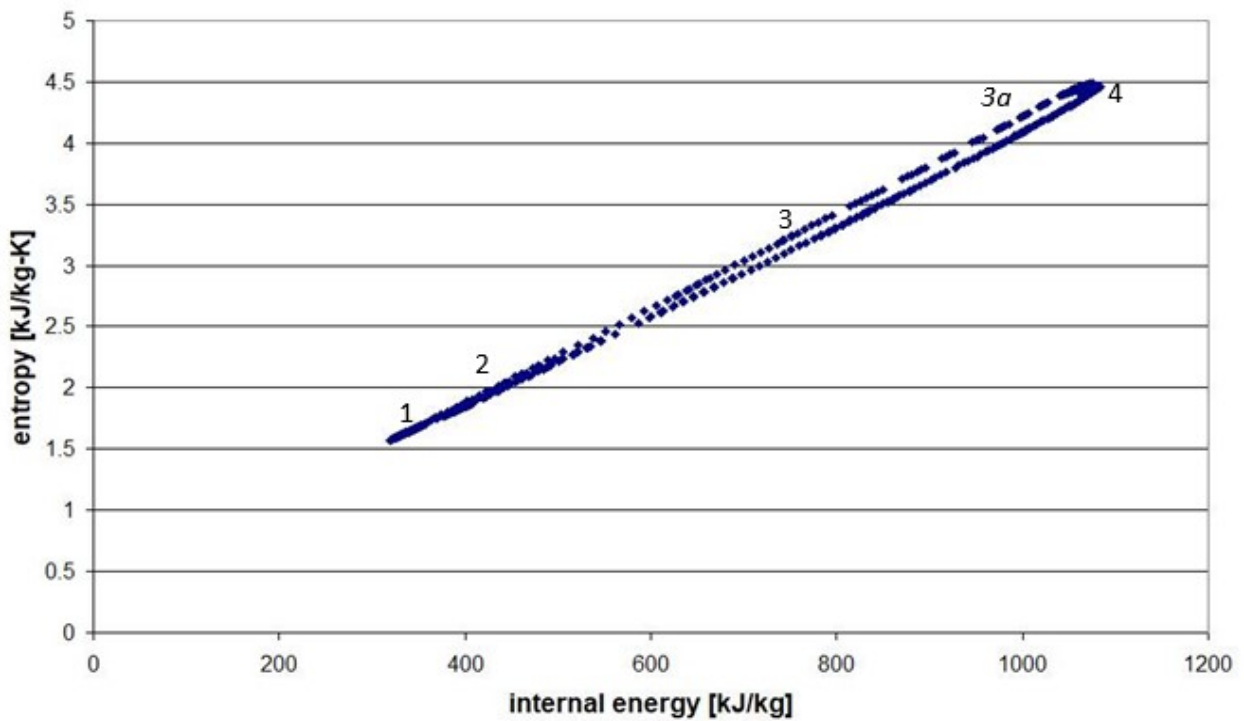

Fig. 2: Affine (non-concave) entropy function revealing a distortion of the short-range interaction

Since enthalpy of the system is dominated by internal energy, a linear enthalpy-entropy compensation effect is also demonstrated.

**Fig. 3** reveals the irreversible hysteresis characteristic of the cycle despite there being no external work input:

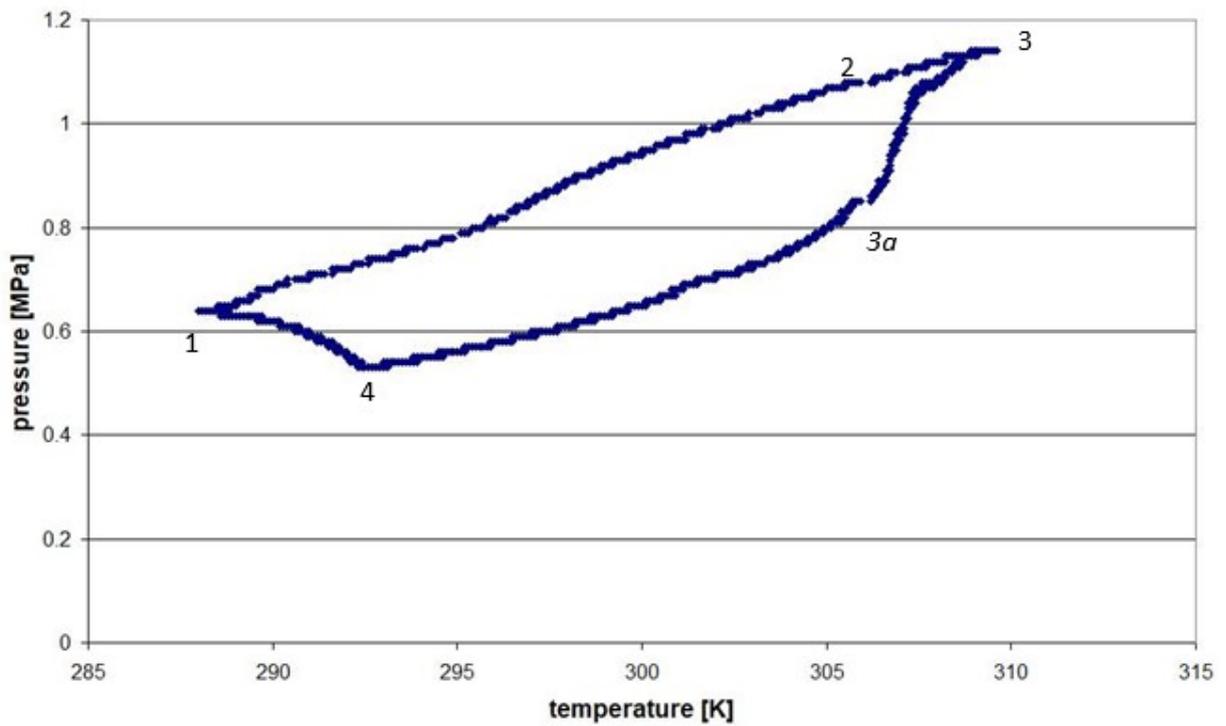

Fig. 3: Hysteresis of 'irreversible' cycle with no work input

**Figs. 4-5** establish the non-equilibrium nature of the cycle since there is no symmetry of energy conservation in terms of the classical thermodynamics properties, as calculated below (ie. an over-unity result is obtained). The cycle is instead considered quasi-thermodynamic.

For a heat engine, net work output is normally given by the bounded area (1-2-3-4-1) of the *P-v* chart, where Stage 4-1 is deducted for isothermal heat rejection and 1-2 is deducted for isentropic compression [28].

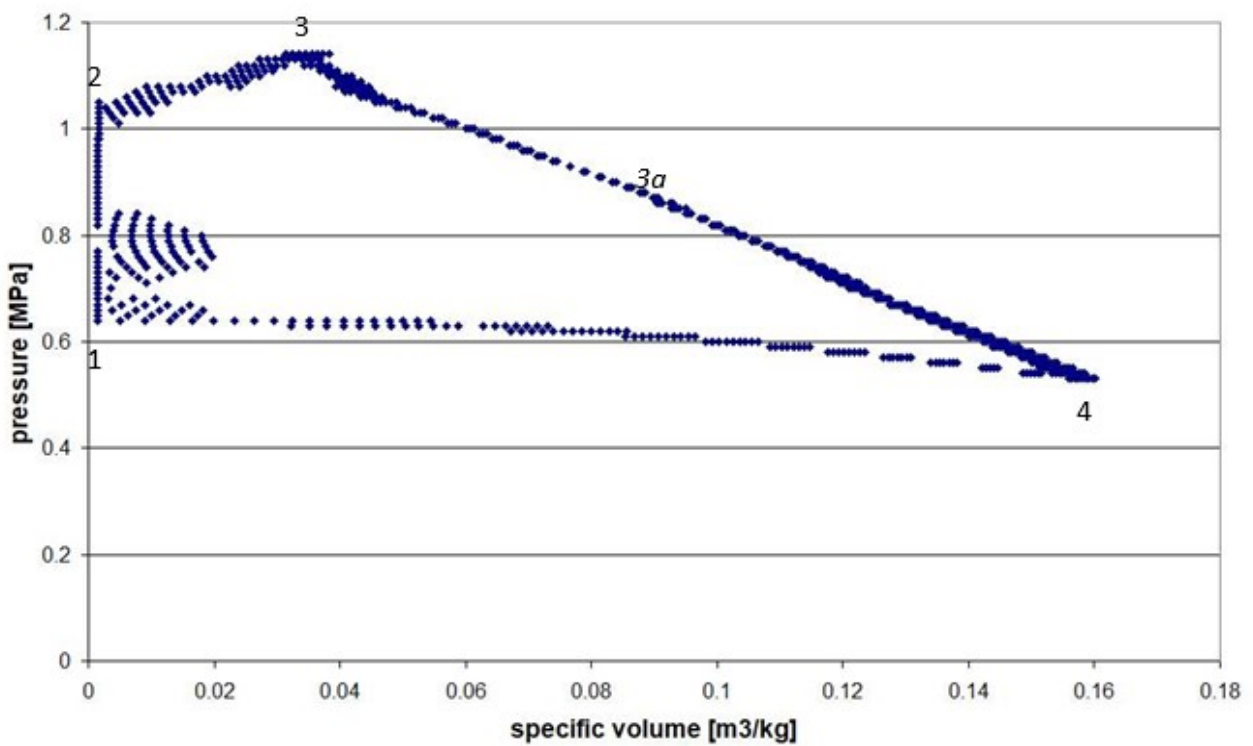

Fig. 4: Phase-change work can be determined from the *P-v* chart

However, in this case with no work input:

>Positive work = bounded area (1-2-3-4-1) + area between 1-4 and the x-axis

>Negative work = area between 4-1 and the x-axis

>**Total = 234 kJ/kg approx.**

Heat input is normally given by the bounded area (1-2-3-4-1) of the *T-s* chart [28].

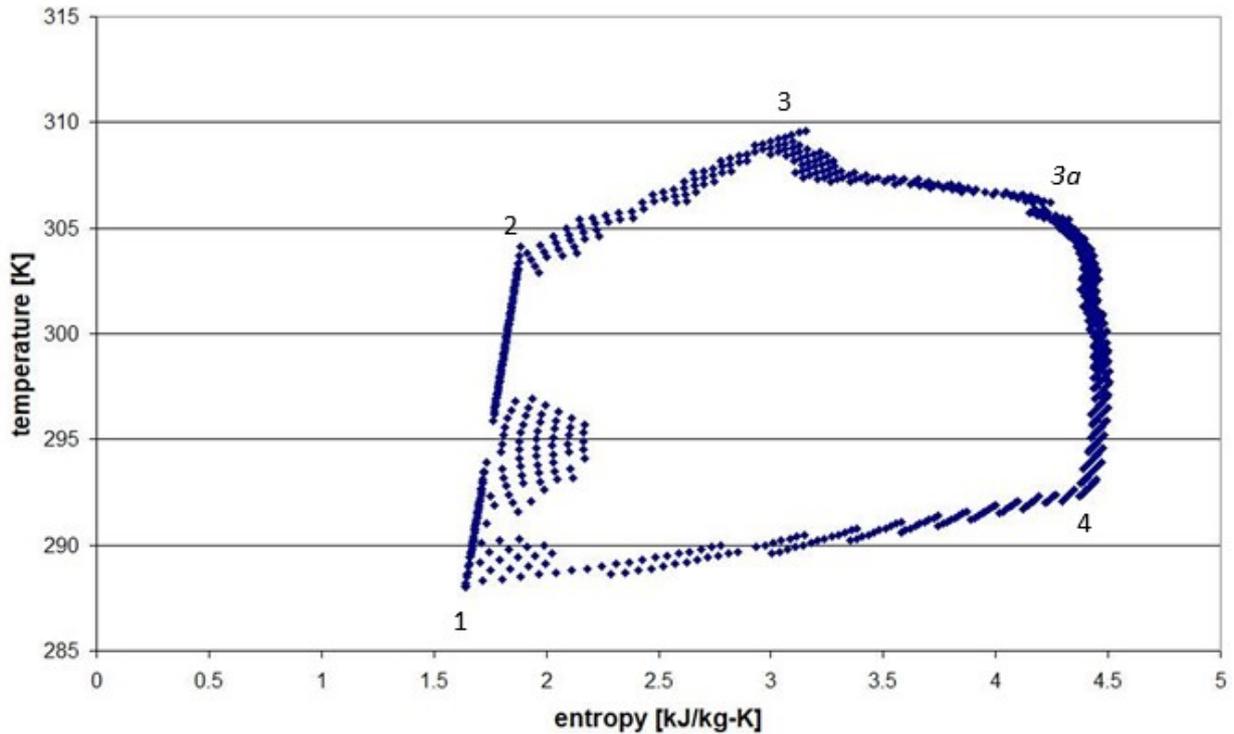

Fig. 5: Heat input can be determined from the *T-s* chart

However, in this case, heat is only supplied during Stages 1-2 and 2-3. Cooling Stage 3-4 represents heat recovery producing positive, expansion work. Cooling Stage 4-1 represents heat recovery producing negative, contraction work. Therefore:

>Heat input = ½ bounded area (1-2-3-4-1)

>**Total = 14.8 kJ/kg approx.**

>The ratio of work output to heat input is therefore **15.8 : 1**.

If the heat recovery stages of 3-4 and 4-1 are treated as isothermal heat rejection and isentropic compression energy penalties, as is the case for a conventional Rankine cycle, then work output and heat input are determined from the bounded areas (1-2-3-4-1). In such a case, the ideal efficiency would be approximately 5% and fall within the Carnot limit. This result allows us to have confidence in the thermodynamic properties calculated by REFPROP and confirms that non-equilibrium, long-range interactions are not unduly interfering in the thermodynamic calculations.

Rather than compare the isochoric process to an ideal process, a more meaningful measure of the usefulness of the process is to compare the useful output with the change in exergy of the system [29]. The ratio of total enthalpy change to the change in exergy (that is the effectiveness of the cycle) gives a similar ratio, approximately **15:1**.

**Fig. 6** reveals an average $1/\sqrt{\rho}$ relationship for the internal energy, associated with expansion and contraction, where the vapour density is below 100 kg/m³. Where the vapour is of higher density, up to the liquid phase transition at Point 2, the internal energy is approximately constant at 400 kJ/kg.

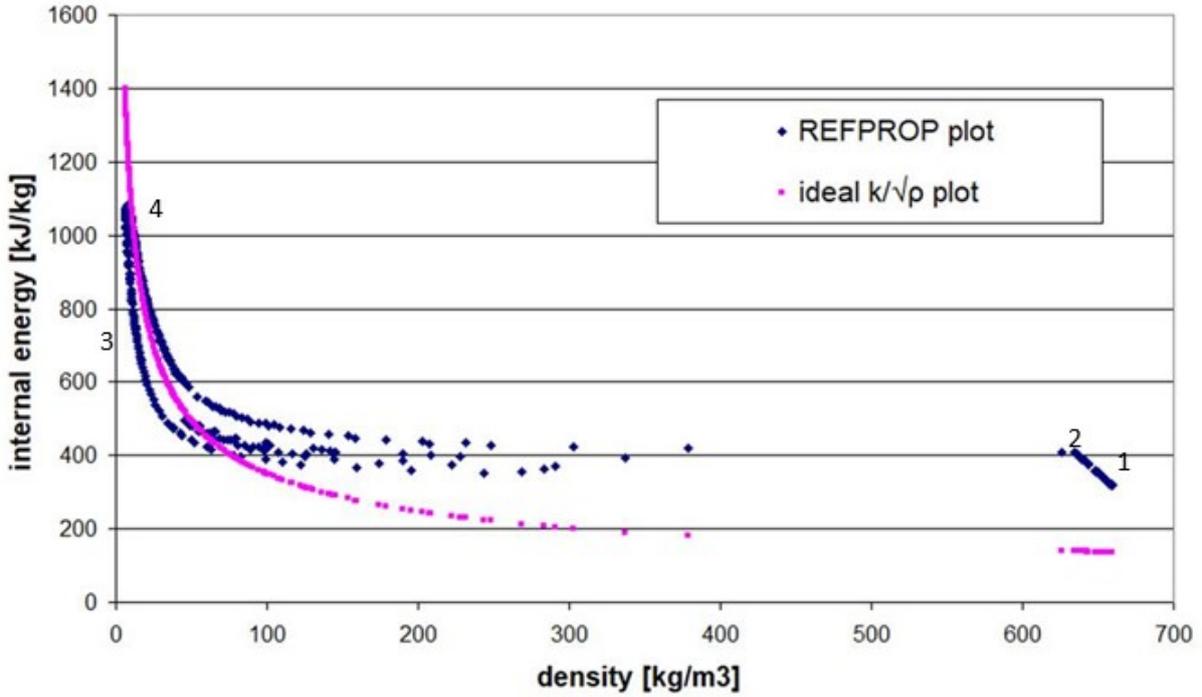

Fig. 6: Internal energy function of density exhibiting an approximate $1/\sqrt{\rho}$ relationship under negative pressure stretching

**Fig. 7** reveals an average $1/\sqrt{N_{guest}}$ relationship where the internal energy exceeds 400 kJ/kg and $N$ is the number of methane molecules hosted by the liquid component of the fluid. These methane molecules are deemed to be hosted by clathrate hydrates, as described later. For lower internal energy values, and down to the liquid phase transition at Point 2, the internal energy is again approximately constant at 400 kJ/kg.

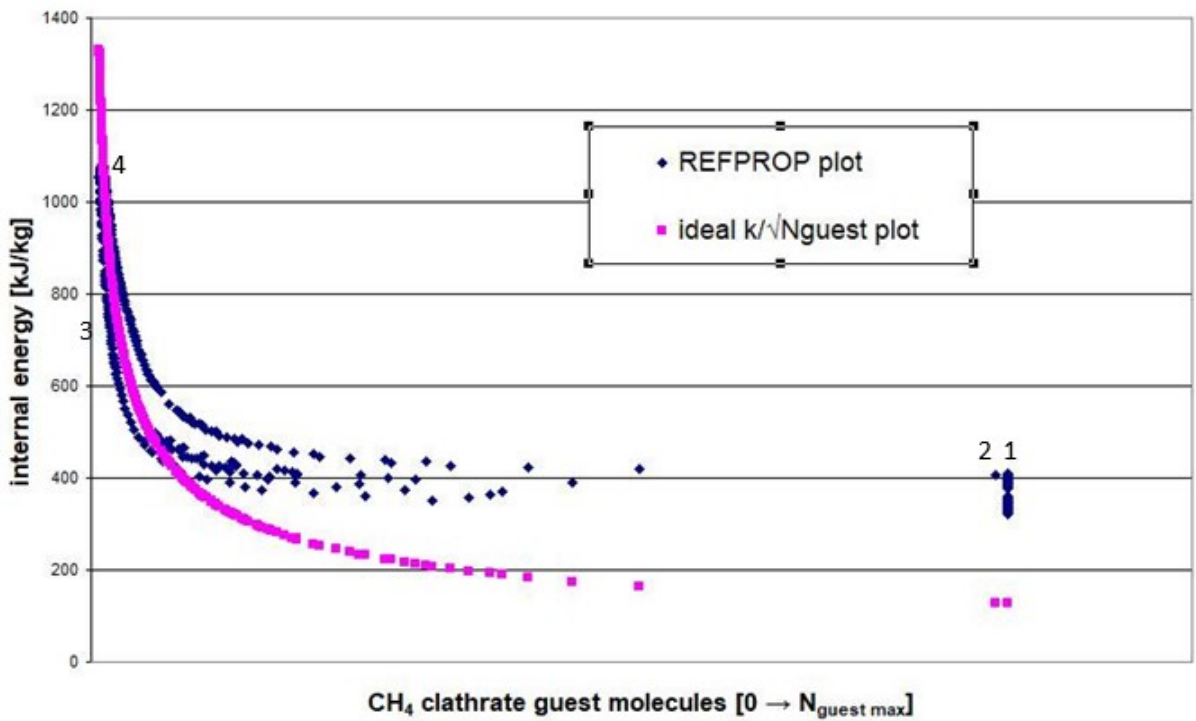

Fig. 7: Internal energy function of $CH_4$ clathrate guest molecules exhibiting an approximate $1/\sqrt{N_{guest}}$ relationship under negative pressure stretching

A comparison of the internal energy values from REFPROP is made with the fundamental thermodynamic relation in the analysis section below. Summary tables of the calculation results are included in Appendix 2. From the inequality between the internal energy and the fundamental thermodynamic relation, a negative excess energy potential is revealed. This is plotted in **Fig. 8**:

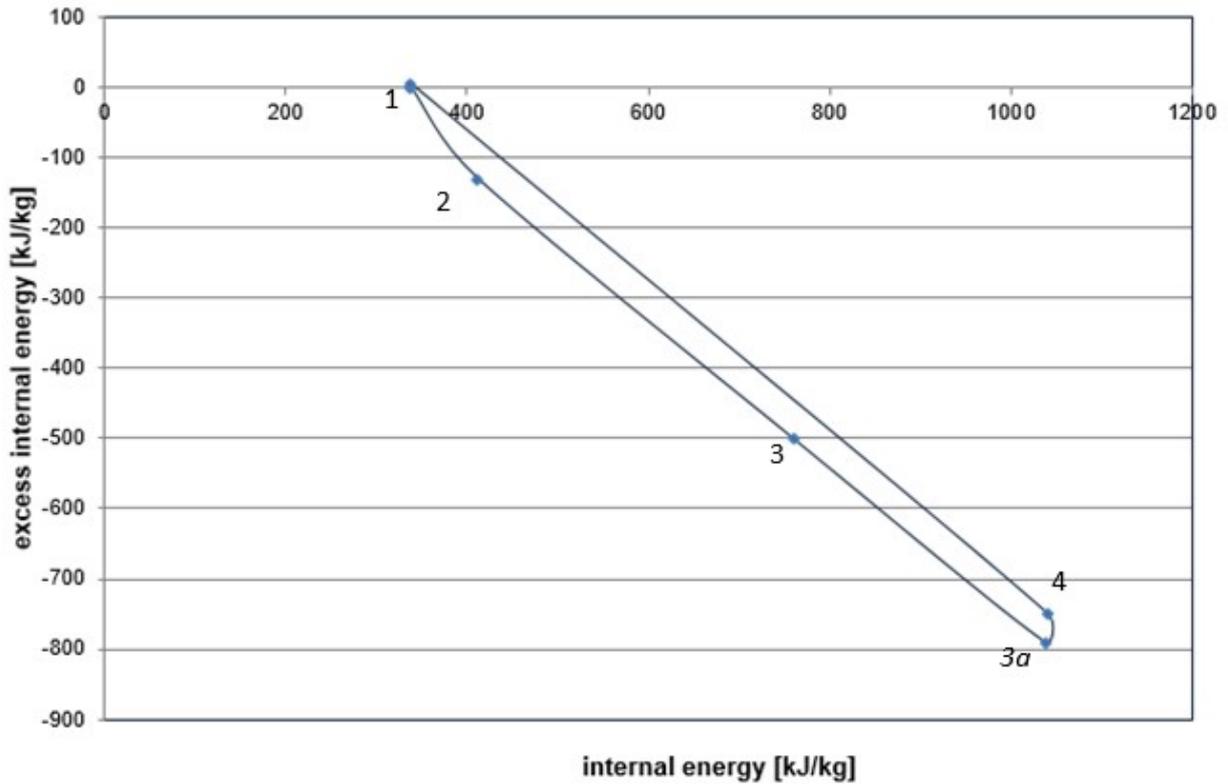

Fig. 8: Excess internal energy cycle resulting from excess thermodynamic potentials

*Establishing a quasi-thermodynamic cycle*

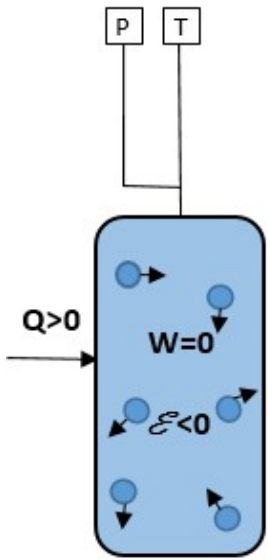

Fig. 9: Thermo-compression

**Stage 1-2** Thermo-compression ($Q$) operates on a sub-cooled liquid such that no mechanical work input ($W$) is required. The stress-strain potential (Helmholz free energy) and the chemical potential (Gibbs free energy) of the polar dielectric fluid remain essentially in balance, such that no $P$-$v$ work term results. A localised cavitation event at 0.6 MPa suggests a liquid-liquid phase transition, or separation, accompanied by limited outgassing and re-absorption of methane. Following this transient phase-splitting, the fundamental thermodynamic relation becomes slightly unbalanced with respect to the thermodynamic potentials although the fluid displays stable properties and behaviour generally consistent with classical thermodynamics.

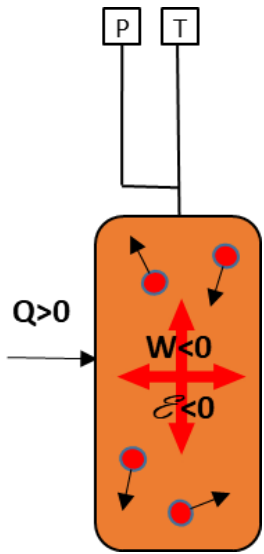

Fig. 10: Endothermic heating

**Stage 2-3** From Point 2, further heating results in an endothermic phase-change process. Divergence of the stress-strain potential and the chemical potential from methane outgassing results in *P-v* phase-change work accompanied by a pressure increase of 0.1 MPa, approx. The fundamental thermodynamic relation is no longer in balance with the thermodynamic potentials. The fluid resembles the metastable false vacuum since the negative pressure fluid maintains its total energy density, as increasing internal energy is balanced by an increasingly negative excess energy potential, $\mathcal{E}$ (after discounting the *P-v* term associated with the walls of the sample vessel).

The system is subject to a long-range interaction, as revealed by the non-additive energies and potentials and the non-concave (affine) entropy function of internal energy. The system is behaving as an analogue of the expanding, false vacuum although the specific heat capacity remains positive throughout this stage. The non-equilibrium fluid behaviour can no longer be described by classical thermodynamics but can be understood in terms of the micro-canonical ensemble of statistical mechanics with constant energy density.

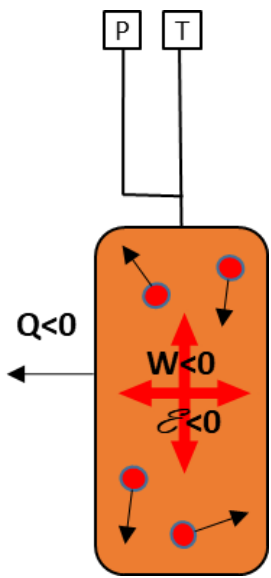

Fig. 11 Endothermic cooling

**Stage 3-3a** Cooling results in further endothermic phase-change expansion work with increasing internal energy. Continued divergence of the stress-strain potential and the chemical potential produces a further *P-v* work term. The fundamental thermodynamic relation is pushed further away from equilibrium with the thermodynamic potentials. The fluid still resembles the metastable, false vacuum as a negative pressure fluid with constant total energy. Increasing internal energy is balanced by an increasingly negative excess energy potential, $\mathcal{E}$ (after discounting the *P-v* term associated with the walls of the sample vessel). The system remains subject to a long-range interaction with non-additive energies and potentials and a non-concave (affine) entropy function of internal energy is displayed. The system continues to be an analogue of an expanding, false vacuum and the specific heat capacity has become negative for this stage. Again, the non-equilibrium fluid behaviour can only be described in terms of the micro-canonical ensemble with constant energy density.

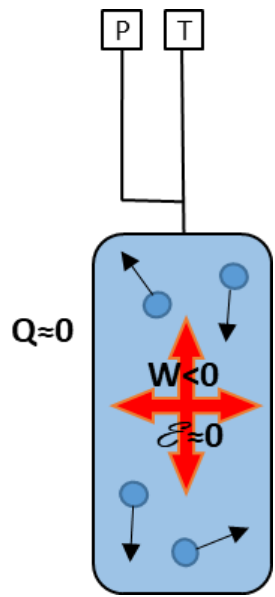

Fig. 12: Isentropic expansion

**Stage 3a-4** Further cooling results in isentropic expansion culminating in the last section of positive *P-v* work. The fluid loses its negative pressure, negative heat capacity characteristics and the analogue false vacuum decays. Both internal energy and negative excess energy potential peak simultaneously and diminish together. By Point 4, the system has excess potential energy of -750 kJ/kg, approx. The system is still behaving as a micro-canonical ensemble but is now entering into an unstable phase of metastability.

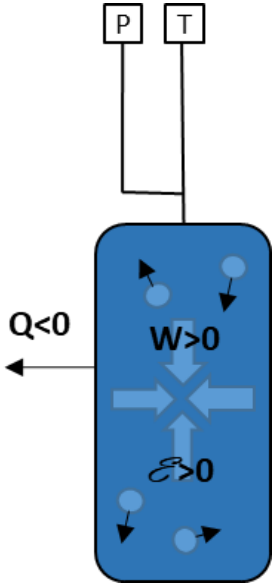

**Stage 4-1** Heat flux is removed which has the effect of pushing the system into an unstable region where a 'non-equivalence of ensembles' condition becomes established. The system displays both micro-canonical (constant energy density) and canonical (almost constant temperature) characteristics. The negative excess energy potential resolves to zero, such that energies become additive and canonical again at Point 1. Internal energy and entropy reduce through a spontaneous relaxation/ nucleation process and the system moves towards a stable liquid phase through the action of an inward-directed, emergent force. A 'non-equivalence of ensembles' phase transition takes place with increasing free energy and pressure. This results in 'negative work' output that condenses and compresses the fluid, returning it to the cycle start at Point 1. The 'negative work' of the emergent force results in -750 kJ/kg, approx. of enthalpy change. The total of the *P-v* work terms across a single cycle is 234 kJ/kg, approx.

Fig. 13: 'Non-equivalence
of ensembles'

Analysis & discussion

The various stages of the cycle are compared with the fundamental thermodynamic relation:

$$dU = Tds - Pdv + \sum_i \mu_i \, dn_i$$

(1)

where *T* is the temperature, *P* is the pressure exerted on the boundary of the system, *v* is the specific volume, *s* is the specific entropy and

$$\sum_i \mu_i \, dn_i$$

is the specific chemical potential (Gibbs Energy).

The summary results included in Appendix 2.

The cycle stages are compared in terms of work transfer and thermodynamic (TD) potentials in **Table 2**:

$$Pv = \sum_i \mu_i \, dn_i - \tfrac{1}{2}\, \sigma_{ij}\, \varepsilon_{ij}$$

(2)

where

$$\tfrac{1}{2}\, \sigma_{ij}\, \varepsilon_{ij}$$

is the specific stress-strain potential (Helmholz Energy) of the negative pressure fluid under tension.

The cycle stages are also compared in terms of heat transfer and thermodynamic potentials in **Table 3**:

$$Ts = 2\left(\sum_i \mu_i \, dn_i\right) - \tfrac{1}{2} \sigma_{ij} \varepsilon_{ij} \tag{3}$$

Returning to the fundamental thermodynamic relation, the excess internal energy is determined in **Table 4**:

$$\text{Excess } dU = \text{excess } Tds - \text{excess } Pdv + \Delta \sum_i \mu_i \, dn_i \tag{4}$$

The largest source of excess energy derives from the negative entropy potentials associated with Stages 2-3 and 3-3a.

The excess negative non-equilibrium potential of expansion is approximately balanced by the excess positive non-equilibrium potential of contraction such that the cycle can return to its start-point equilibrium with no external work input. When returning to the liquid phase (around Point 1) the excess non-equilibrium potentials are resolved.

The results in **Table 3** suggest that an analogue of the false vacuum comes into existence during the positive, expansion stages of the cycle. The excess potential energy evolved is equal-and-opposite to the increase in internal energy. Expressed in terms of an equilibrium system, the excess potential revealed through the fundamental thermodynamic relation may also be considered as 'replica energy' ($\mathscr{E}$), since the potential appears to derive from statistical replicas of the system that differ only in phase, ie. with respect to configuration and velocity [13]. However, in this case most of the phase-changes across the cycle occur under non-equilibrium conditions.

In order to establish a false vacuum condition, that is one with 'non-additive' behaviour [30], an attempt is made to extend the statistical mechanics of 'small systems' to open, metastable, supersaturated gaseous systems that are close to the gas-liquid equilibrium transition point [31]. Non-additive energies combined with non-concave entropy functions of internal energy [6,18] represent long-range distortions to the basic short-range interactions of thermodynamic systems. These distortions, or violations, are at the origin of ensemble non-equivalence, negative specific heat and ergodicity breaking.

**Fig.2** reveals a fluid moving through various states of metastable non-equilibrium. However, internal energy *u* is normally envisaged as the 'random molecular energy' of a closed system under equilibrium. Within the kinetic theory of fluids, the only forms of random molecular energy that can change are the kinetic energy of the molecules and the potential energy due to molecular forces. But when the system is out of equilibrium, the internal energy can include macroscopic mechanical forms, both kinetic and potential [29].

Since the Coulomb force and the gravitational force are described in 3-dimensional space by an inverse-square law, it seems possible that the polar dielectric fluid may act as an analogue of a self-gravitating system. However, such an analogy should be frustrated by ionic charge interactions which are expected to make a significant contribution to the total potential energy, and thus the internal energy interactions between fluid molecules, producing an electrostatic screening effect. Within statistical physics, gravity is an unscreened, long-range interaction that produces non-additivity, whereas the Coulomb force is usually described as a screened, short-range interaction.

Gebbie *et al* [32] cite experimental results revealing that ionic liquids can have remarkably long-range interactions that appear to be electrostatic in origin. In one case it was suggested that less than 0.1% of the total number of ions were fully dissociated and independently contributed to electrostatic screening.

It was also predicted that ionic liquids with higher dielectric permittivity would exhibit higher degrees of ionic dissociation resulting in shorter electrostatic screening lengths.

Permittivity is a material property that affects the Coulomb force between point charges in the fluid. Combining Einstein's mass-energy equivalence formula with Maxwell's electromagnetic wave equation establishes a relationship between the total energy of a system and relative permittivity/ relative permeability. A change in total energy of a dielectric fluid of fixed mass is associated with a change in relative permittivity/ relative permeability as determined by:

$$E_1 = E/\sqrt{((\varepsilon_0/\varepsilon_s)(\mu_0/\mu_s))} \qquad (5)$$

where $E_1/E$ is the relative change in total energy, $\varepsilon_0$ is the permittivity of free space; $\mu_0$ is the permeability of free space; $\varepsilon_s$ is the static permittivity of the fluid; $\mu_s$ is the static permeability of the fluid.

For the micro-canonical ensemble, where $E_1 = E$, the effect of any change in relative permittivity must be exactly matched by an equal-and-opposite change in the effect of relative permeability, ie. the product $\varepsilon_s \cdot \mu_s$ results in a constant. As the fluid undergoes its liquid-vapour phase transition, we should expect the value of relative permittivity to reduce significantly (so increasing the electrostatic Coulomb potential), with such a change mirrored by an increase in relative permeability (so reducing the magnetostatic Coulomb potential). Such a mechanism may indicate the presence of a 'Coulomb fluid', or 'magnetolyte' in which weakly dissociated ions of water are correlated through hydrogen-bonded chains [33].

The model proposed by Gebbie *et al* is for a small thermally excited population of 'free ions' in equilibrium with a strongly correlated ionic network where each charge is neutralised by the sum of neighbouring charges, ie. a dielectric solvent [32]. It is proposed that the action of heat and negative pressure on our polar dielectric fluid formulation is responsible for the appearance of similarly long-ranged 'free ion' interactions, as described below.

For most physical systems the non-additivity of energy interactions implies non-extensivity resulting from long-range interactions. For the case being considered, these long-range interactions are seen to distort the short-range thermodynamics to produce a 'non-equivalence of ensembles' [18].

**Fig. 2** also underpins an enthalpy-entropy compensation effect which can indicate solvent reorganisation leading to supramolecular encapsulation by water [34]. Guillot and Giussani [35] previously examined the solubitity of methane in water through the application of molecular dynamics and also found that the pair distribution function between solute and solvent enabled the formation of clathrate-type cages around the solute.

The enthalpy-entropy compensation effect is attributed to the entropy change of solvent reorganisation cancelling out the associated enthalpy change in the contributions made to the Gibbs free energy [36]. The free energy change is then directly related to the non-compensating part of the entropy change that arises simply from the exclusion zone that has to exist around the cavity.

**Fig. 3** illustrates an irreversible process taking place. Any process involving non-equilibrium states is irreversible, as is a thermodynamic process requiring external work to restore the working fluid to its initial condition [28]. Since there is no external work input for the case considered, and the isochoric cycle is performed over an 11-hour period, it seems reasonable to expect the cycle to be approximately internally reversible. However, the path described clearly displays the irreversible characteristic of hysteresis that hints at a non-equilibrium source of work.

Far-from-equilibrium systems are not generally characterised by an extremum principle, eg. a tendency to minimise energy or maximise entropy, thus they becomes more unstable and fluctuations can lead to other, more stable states [37]. These new states are often expressed through higher degrees of organisation that involve concepts such as 'dissipative structures' and 'self-organisation' [12]. In open systems, dissipative structures can be maintained indefinitely through a flow of matter and energy. Irreversible processes are not usually governed by global extremum principles because a description of their evolution requires differential equations which are not self-adjoint. However, local extremum principles, ie. metastable conditions, can be used for local solutions [38].

**Figs. 4-5**, together with the associated cycle efficiency calculations, do not produce a symmetry of energy conservation in terms of classical thermodynamics (or 'thermostatics' [39]). It is thereby, again, evident that we are dealing with a quasi-thermodynamic cycle operating under non-equilibrium conditions. The non-equilibrium potential energy component of the cycle can be established from the symmetry of energy conservation, ie. the sum of equilibrium and non-equilibrium energies, joint and several, must equal zero when proceeding through a full cycle.

Non-equilibrium systems can be broadly classified as near-equilibrium systems, in which there is a linear relation between forces and flows (the linear regime), and far-from-equilibrium systems, in which the relationship between forces and flows is nonlinear (the nonlinear regime). In the near-equilibrium, linear regime, Onsager reciprocal relations apply. In the far-from-equilibrium, nonlinear regime, spontaneous self-organisation and emergent, dissipative structures are evident [10].

**Figs. 6-7** The internal energy functions (the $1/\sqrt{\rho}$ relationship and $1/\sqrt{N_{guest}}$ relationship) are essentially the same revealing the density of the fluid to be directly related to the number of methane molecules within the liquid or liquid component of the vapour. The ideal $1/\sqrt{\rho}$ and $1/\sqrt{N_{guest}}$ relationships are straddled by the irreversible expansion and contraction curves when above 400 kJ/kg.

The mirroring of internal energy with excess negative potential energy, combined with non-additivity associated with non-equilibrium conditions, reveals the existence of negative pressure, or internal tension. Our analysis above also confirms the excess potential energy to be dominated by negative entropy. The empirical results further reveal a decrease in phase density of the polar dielectric fluid with increasing negative pressure conditions, or stretching, favourable to the formation of structural cavities and cages [36,40], ie. lower density, inclusion compound structures [21,41]. The presence of methane and water in the polar dielectric fluid formulation allows us to reasonably infer that these inclusion compounds are methane clathrates, behaving as soft supramolecular materials, responsive to external stimuli and readily converted from one structure to another [42].

In our results, reciprocal relations briefly occur before the cavitation event in Stage 1-2, for entropy of 1.6 kJ/kg-K and internal energy of 400 kJ/kg. C*onstant* entropy has an equal and opposite cross-effect of *constant* internal energy with the relations taking place at a constant density of 650 kg/m$^3$. Since no flows of matter or total energy are associated with this condition, it should not be possible for dissipative clathrate structures to be sustained through this section of the cycle. This could indicate the onset of a re-organisation of the water molecules from one clathrate structure to another with a transient outgassing of methane occurring during the transition.

Continued heat flow (above 400 kJ/kg of internal energy) pushes the system far-from-equilibrium where even lower density clathrate structures can form. A more definite liquid-gas phase separation, revealed through a more pronounced methane outgassing, produces *P-v* work (with an expansion ratio of 100:1) whilst also facilitating the formation of these lower entropy, guest-free, clathrate structures [41,43,44]. This micro-canonical phase-change would involve fragmentation of the system into a spatially inhomogeneous distribution of various regions with different densities and phases [8]. The high negative internal pressure necessary for this process is created through the interaction of the inhibitor solvent with the water structures [24], which also prevents any liquid-solid phase separation, or agglomeration.

With the internal energy function displaying both a $1/\sqrt{\rho}$ and $1/\sqrt{N_{guest}}$ relationship, a long-range interaction is evident for the far-from-equilibrium conditions. Employing the model proposed by Gebbie *et al* [32], the requirement for long-range 'free ion' interaction appears satisfied by the attraction between guest-free clathrates and their former guests through long-range van der Waals forces [41,43]. The rate of methane outgassing and re-absorption satisfies the predictions of the fluctuation-dissipation theorem where the size of the fluctuations scales as:

$$\Delta E/E \sim \sqrt{N_{guest}} \tag{6}$$

$N_{guest}$ is the number of guest methane molecules within the liquid component of the fluid. The fluctuations of the system are related to the ability of the system to dissipate or absorb energy [45]. The rate of methane outgassing reduces as a power function, from a maximum to a minimum, as the number of guest-free clathrate cages multiplies and the mean distance between guests and hosts increases (ie. the dissipation potential reduces). Conversely, the rate of methane re-absorption increases as the same power function, from a minimum to a maximum, as clathrate cages once again play host to methane molecules and the mean separation distance reduces. The re-absorption process is examined further below.

Moving even further away from equilibrium, with still higher negative pressures, we are able to establish negative heat capacity during Stage 3-3a where outgassing of the residual methane occurs. Negative specific heat arises from system fragmentation [8] just as it arises in core-halo structures [46]. It appears that the entropic and structural inertia of the guest-free clathrates gives rise to a 'self-gravitating' system where cooling of a system having negative specific heat gives rise to fluid expansion with increasing kinetic energy [47]. Where a negative pressure, self-gravitating system expands, its increasing kinetic energy will be mirrored by an increasingly negative potential energy [9].

If the entropy-flow leaving the liquid component is larger than the one entering, the liquid evacuates its entropy by an irreversible process that creates internal order, or structure [46]. This would lead to the formation of even lower density, guest-free clathrate structures. By Point 3a the flow of methane from liquid to gas components is complete. However, there immediately follows one final stage of *P-v* expansion. From Point 3a to 4, isentropic expansion associated with a limited liquid-gas phase-change of the inhibitor solvent is driven by the still increasing negative pressure of the cooling expansion.

At Point 4 heat flux is removed and a 'non-equivalence of ensembles' condition becomes established. Since the system is no longer driven by flows of matter and energy, it becomes unstable such that internal entropy is able to rebalance through a relaxation process, re-establishing a canonical ensemble in the liquid phase [48]. Cooling expansion has reduced the kinetic energy of the gas molecules and, consequently, the large potential wells of guest-free clathrates in the liquid begin to recapture their former guests from the expanded gas; a process similar to gravitational clustering in core-halo configurations, as described by Padmanabhan [7]. The rate of re-absorption, initiated by large inhomogeneous fluctuations [8], increases as the gas volume collapses and the increase in $N_{guest}$ molecules drives the energy dissipation potential of the system [45].

**Figs. 6-7** describe internal energy functions in terms of inverse square root relationships of both density and the number of clathrate-hosted methane molecules. The clathrates and inhibitor solvent are essentially composed of tetrahedral structures, giving a co-ordination number of 4. In terms of the second moment approximation of tight binding theory, where the cohesive energy varies as the square root of the coordination number [49], the potential energy function is a ($-1/\sqrt{r}$) function, such that $\alpha = -1/2$ for an *internal* non-central potential, ie. the gas envelopes the liquid solvent/ guest-free clathrate clusters in a large-series approximation of a core-halo structure. However, expressed in terms of an *external*, central potential [50], $\alpha$ again becomes -2 such that $U = K = -V$ is confirmed. As an analogue of a self-gravitating system, the separation of methane guests from clathrate hosts in an ionic inhibitor fluid establishes a long-range van der Waals interaction; the result of thermally-driven, negative pressure 'stretching' of the polar dielectric fluid. In this case the separated clathrate hosts and guests are proposed as the 'free ions' identified by Gebbie *et al* [32].

**Fig. 8** describes the characteristics of a linear oscillator where changes in negative potential energy and internal energy vary in an approximate 1:1 relationship, after discounting the *P-v* term associated with the walls of the sample vessel. The virial theorem states that for gravitationally bound objects:

$$2<K> + <V> = 0 \quad \text{or} \quad <K> = -<V>/2 \tag{7}$$

A more general case applies for a system of particles interacting with a potential in the form $V = 1/r^\alpha$ [51] for which the expression becomes:

$$2\langle K \rangle = \alpha \langle V \rangle \qquad (8)$$

For a linear oscillator responding to an analogue gravity potential $\alpha$ is taken as -2 [52] which gives:

$$\langle K \rangle = -\langle V \rangle \qquad (9)$$

Conclusion

Thermal manipulation of a polar dielectric fluid under isochoric conditions establishes a negative pressure, metastable system analogous to the false vacuum. As with the false vacuum, the system seeks to maintain a constant energy density such that increasing kinetic energy is mirrored by an increasing negative energy potential. Evidence for the existence of this negative energy potential is established through the non-additivity of thermodynamic potentials within the fundamental thermodynamic relation and the non-concave entropy function of internal energy. These canonical violations can be recovered in terms of the micro-canonical ensemble of statistical mechanics. However, since the ultimate source of the negative energy potential is unknown, it is not clear that the system is strictly isolated.

Changes in fluid permittivity affect the Coulomb force. From our findings, the short-range screening effect is neutralised in a 'dielectric solvent' inhibitor such that a long-range van der Waals interaction can emerge. This is described by a $1/\sqrt{\rho}$ or $1/\sqrt{N_{guest}}$ relationship hence the excess negative energy potential corresponds to a $-1/\sqrt{r}$ relationship (or $-1/r^2$ for an external, central potential). This unusual result appears to be associated with ineffective dissociation of inhibitor solvent molecules which reduces electrostatic screening effects. A long-range interaction can then established from the van der Waals attraction between guest-free clathrate hosts and their former guests. An analogue of a self-gravitating system is thereby established and shown to be consistent with the virial theorem.

Within the thermodynamic cycle described, the long-range interaction manifests as work output operating in two directions, positive expansion and negative contraction. A large proportion of expansion work derives from a cooling process. Negative contraction work results from the instability associated with the 'non-equivalence of ensembles' whilst the negative heat capacity responsible for cooling-expansion is a phenomenon associated with classic fluctuation theory.

Once negative excess energy potentials are incorporated into the fundamental thermodynamic relation, a symmetry of energy conservation is established for the complete 'quasi-thermodynamic' cycle. The excess negative energy potential of the attractive van der Waals interaction is revealed to be distinct from the heat and work input of conventional thermodynamic cycles with the resultant emergent force contributing additionally to overall work output and effectiveness of the quasi-thermodynamic cycle.

Appendix 1

The recorded data and REFPROP calculated properties are presented in the tables and figures below:

Table 1: Recorded and calculated thermodynamic properties of a typical cycle

|  | Temperature T (K) | Pressure P (MPa) | specific volume v (m3/kg) |
|---|---|---|---|
| Point 1 | 288.0 | 0.64 | 0.0015 |
| Point 2 | 304.6 | 1.05 | 0.0015 |
| Point 3 | 309.6 | 1.14 | 0.0382 |
| Point 3a | 306.2 | 0.85 | 0.0951 |
| Point 4 | 292.3 | 0.53 | 0.1596 |

|  | internal energy u (kJ/kg) | Gibbs energy G (kJ/kg) | Helmholz energy F (kJ/kg) |
|---|---|---|---|
| Point 1 | 338 | -133 | -134 |
| Point 2 | 412 | -161 | -162 |
| Point 3 | 760 | -173 | -216 |
| Point 3a | 1039 | -179 | -260 |
| Point 4 | 1040 | -159 | -242 |

|  | entropy s (kJ/kg-K) | exergy b (kJ/kg) | enthalpy h (kJ/kg) |
|---|---|---|---|
| Point 1 | 1.64 | 863 | 340 |
| Point 2 | 1.89 | 865 | 414 |
| Point 3 | 3.15 | 843 | 803 |
| Point 3a | 4.24 | 798 | 1120 |
| Point 4 | 4.38 | 757 | 1137 |

Appendix 2

A comparison is made between the internal energy results and the fundamental thermodynamic relation.

Table 2: Work vs. thermodynamic relations

|  | 1 | 2 | 1 - 2 |  |
|---|---|---|---|---|
|  | Pv from TD potentials kJ/kg | Pv from REFPROP kJ/kg | Excess Pv kJ/kg | ∑ Excess Pv kJ/kg |
| Stage 1-2 | 0 | 0 | 0 | 0 |
| Stage 2-3 | 42 | 40 | 2 | 2 |
| Stage 3-3a | 38 | 57 | -19 | -17 |
| Stage 3a-4 | 2 | 44 | -42 | -59 |
| Stage 4-1 | -82 | -92 | 10 | -49 |

Table 3: Heat vs. thermodynamic relations

|  | 1 | 2 | 1 - 2 |  |
|---|---|---|---|---|
|  | Ts from TD potentials kJ/kg | Ts from REFPROP kJ/kg | Excess Ts kJ/kg | ∑ Excess Ts kJ/kg |
| Stage 1-2 | -28 | 74 | -102 | -102 |
| Stage 2-3 | 30 | 387 | -357 | 255 |
| Stage 3-3a | 32 | 336 | -304 | 559 |
| Stage 3a-4 | 22 | 42 | -20 | 579 |
| Stage 4-1 | -56 | -795 | 739 | -160 |

Table 4: Summary of excess thermodynamic potentials

|  | 1 | 2 | 3 | 1 – 2 + 3 |  |
|---|---|---|---|---|---|
|  | Excess Ts kJ/kg | Excess Pv kJ/kg | Δ Gibbs kJ/kg | Excess u kJ/kg | ∑ Excess u kJ/kg |
| Stage 1-2 | -102 | 0 | -28 | -130 | -130 |
| Stage 2-3 | -357 | 2 | -12 | -371 | -501 |
| Stage 3-3a | -304 | -19 | -6 | -291 | -792 |
| Stage 3a-4 | -20 | -42 | 20 | 42 | -750 |
| Stage 4-1 | 739 | 10 | 26 | 755 | 5 |